\documentclass[longauth,manuscript]{aa}
\usepackage{color,epsfig,graphics,graphicx}

\usepackage{natbib,ulem}
\usepackage{multirow}
\usepackage{natbib,ulem,lineno}
\usepackage{graphicx}
\usepackage{subcaption}
\usepackage{txfonts}

\definecolor{brown}{rgb}{0.6,0.4,0.2}
\definecolor{purple}{rgb}{0.5,0,0.5}

\def\msun{$M_{\odot}$}
\def\one{{\,\sc i}}
\def\two{{\,\sc ii}}
\def\three{{\,\sc iii}}
\def\four{{\,\sc iv}}
\def\five{{\,\sc v}}
\newcommand{\iso}[2]{\ensuremath{^{#1}\rm{#2}}}
\def\nifs{\iso{56}Ni}

\newcommand{\kms}{km\,s$^{-1}$}

\newcommand{\Ha}{H\,$\alpha$}

\newcommand{\mic}{$\mu$m}

\newcommand{\new}{\textcolor{black}}

\newcommand{\newb}{\textcolor{blue}}

\begin{document}

    \title{Near-Infrared Spectroscopy and Detection of Carbon Monoxide in the Type II Supernova SN\,2023ixf}

    \author{Seong Hyun Park\inst{1}, Jeonghee Rho\inst{2,1}, Sung-Chul Yoon\inst{1}, Jeniveve Pearson\inst{3}, Manisha Shrestha\inst{3}, Samaporn Tinyanont\inst{4,5}, T. R. Geballe\inst{6}, Ryan J. Foley\inst{5}, Aravind P. Ravi\inst{7}, Jennifer Andrews\inst{6}, David J. Sand\inst{3}, K. Azalee Bostroem\inst{3}\fnmsep\thanks{LSST-DA Catalyst Fellow}, Chris Ashall\inst{8}, Peter Hoeflich\inst{9}, Stefano Valenti\inst{7}, Yize Dong\inst{7}, Nicolas Meza Retamal\inst{7}, Emily Hoang\inst{7}, Darshana Mehta\inst{7}, D. Andrew Howell\inst{10,11}, Joseph R. Farah\inst{10,11}, Giacomo Terreran\inst{12,10}, Estefania Padilla Gonzalez\inst{13,10}, Moira Andrews\inst{10,11}, Megan Newsome\inst{10,11}, Melissa Shahbandeh\inst{13}, Nathan Smith\inst{3}, Jae Hwan Kang\inst{14}, Nick Suntzeff\inst{15}, Eddie Baron\inst{16,17,18}, Kyle Medler\inst{8}, Tyco Mera Evans\inst{9}, James M. DerKacy\inst{13,8}, Conor Larison\inst{19}, Llu\'is Galbany\inst{20,21}, and
    Wynn Jacobson-Gal\'an\inst{14}
        }

    \institute{Department of Physics and Astronomy, Seoul National University, Gwanak-ro 1, Gwanak-gu, Seoul, 08826, South Korea\\
              \email{rogersh0125@snu.ac.kr; scyoon@snu.ac.kr} %1
         \and
          SETI Institute, 339 Bernardo Ave., Ste. 200, Mountain View, CA 94043, USA
              \email{jrho@seti.org} %2
         \and
          Steward Observatory, University of Arizona, 933 North Cherry Avenue, Tucson, AZ 85721-0065, USA
         \and
          National Astronomical Research Institute of Thailand, 260 Moo 4, Donkaew, Maerim, Chiang Mai, 50180, Thailand
         \and
          Department of Astronomy and Astrophysics, University of California, Santa Cruz, CA 95064, USA
         \and
          Gemini Observatory/NSF's National Optical-Infrared Astronomy Research Laboratory, 670 N. Aohoku Place, Hilo, HI, 96720, USA
              \email{tom.geballe@noirlab.edu}
         \and
          Department of Physics and Astronomy, University of California, 1 Shields Avenue, Davis, CA 95616-5270, USA
         \and
          Department of Physics, Virginia Tech, Blacksburg, VA 24061, USA
         \and
          Florida State University, Tallahassee, FL 32309, USA
         \and
          Las Cumbres Observatory, 6740 Cortona Drive, Suite 102, Goleta, CA 93117-5575, USA
         \and
          Department of Physics, University of California, Santa Barbara, CA 93106-9530, USA
         \and
          Adler Planetarium, 1300 S. DuSable Lake Shore Dr., Chicago, IL 60605, USA
         \and
          Space Telescope Science Institute, 3700 San Martin Drive, Baltimore, MD 21218-2410, USA
         \and
          California Institute of Technology, Pasadena, CA 91125, USA
         \and
          George P. and Cynthia Woods Mitchell Institute for Fundamental Physics and Astronomy, Department of Physics and Astronomy, Texas A\&M University, College Station, TX 77843, USA
         \and
          Planetary Science Institute, 1700 East Fort Lowell Road, Suite 106, Tucson, AZ 85719-2395, USA
         \and
          Hamburger Sternwarte, Gojenbergsweg 112, D-21029 Hamburg, Germany
         \and
          Homer L. Dodge Department of Physics and Astronomy, University of Oklahoma,  Norman, OK 73019-2061, USA
         \and
          Rutgers Department of Physics and Astronomy, 136 Frelinghuysen Rd, Piscataway, NJ 08854, USA
         \and
          Institute of Space Sciences (ICE-CSIC), Campus UAB, Carrer de Can Magrans, s/n, E-08193 Barcelona, Spain
         \and
          Institut d'Estudis Espacials de Catalunya (IEEC), 08860 Castelldefels (Barcelona), Spain
             }

\abstract
 {Core-collapse supernovae (CCSNe) may contribute a significant amount of dust in the early universe. Freshly formed coolant molecules (e.g., CO) and warm dust can be found in CCSNe as early as $\sim$100~d after the explosion, allowing the study of their evolution with time series observations.}
 {In the Type II SN\,2023ixf, we aim to investigate the temporal evolution of the temperature, velocity, and mass of CO and compare them with other CCSNe, exploring their implications for the dust formation in CCSNe. From observations of velocity profiles of lines of other species (e.g., H and He), we also aim to characterize and understand the interaction of the SN ejecta with preexisting circumstellar material (CSM).}
 {We present a time series of 16 near-infrared spectra of SN\,2023ixf from 9 to 307~d, taken with multiple instruments: Gemini/GNIRS, Keck/NIRES, IRTF/SpeX, and MMT/MMIRS. }
 {The early ($t\lesssim70$~d) spectra indicate interaction between the expanding ejecta and nearby CSM. At $t\lesssim20$~d, intermediate-width line profiles corresponding to the ejecta-wind interaction are superposed on evolving broad P Cygni profiles. We find intermediate-width and narrow lines in the spectra until $t\lesssim70$~d, which suggest continued CSM interaction. We also observe and discuss high-velocity absorption features in H~$\alpha$ and H~$\beta$ line profiles formed by CSM interaction. The spectra contain CO first overtone emission between 199 and 307~d after the explosion. We model the CO emission and find the CO to have a higher velocity (3000-3500 \kms) than that in Type II-pec SN\,1987A (1800 - 2000 \kms) during similar phases ($t=199-307$~d) and a comparable CO temperature to SN\,1987A.  A flattened continuum at wavelengths greater than 1.5~\mic\ accompanies the CO emission, \new{suggesting that the warm dust is likely formed in the ejecta. The warm dust masses are estimated to be on the order of $\sim$10$^{-5}$~\msun.}}
 {}
%{We find signs of early CSM interaction in SN\,2023ixf at optical and near-infrared wavelengths. We also find that the temporal evolution of CO in SN\,2023ixf differs from that of previously reported Type II-pec SN\,1987A.}

\titlerunning{NIR spectroscopy of SN\,2023ixf}
\authorrunning{S. H. Park et al.} 
\maketitle
\section{Introduction}
SN\,2023ixf, located in the M101 galaxy, is the nearest \citep[$d$ = 6.85 $\pm$ 0.15 Mpc;][]{riess22} Type II supernova (SN) in several decades and is one of the best-studied SNe since its discovery. SN\,2023ixf was
reported by \citet{itagaki23} on 2023 May 19 at 21:42 UT and
was spectroscopically classified by \citet{perley23ixftype}.
The estimated explosion time is 2023 May 18 18:00:00 \citep[MJD 60082.75 $\pm$ 0.10 ;][]{hosseinzadeh23}.  

The progenitor of SN\,2023ixf identified from pre-explosion imaging was a red supergiant \citep[RSG; e.g.,][]{jenson23}. Based on comparison with a shock cooling model, \citet{hosseinzadeh23} estimated a progenitor radius of 410 $R_\odot$. The progenitor was obscured by a dusty wind with a mass loss rate of $\dot{M}$ = $10^{-6}-10^{-4}$ \msun~yr$^{-1}$ estimated by the pre-explosion imaging and the variability analysis of the progenitor \citep{jenson23,kilpatrick23,soraisam23,neustadt24}. The estimated pre-explosion luminosity of SN\,2023ixf is $\log{(L/L_{\odot})}$~$\sim$~$4.7-5.2$, which corresponds to a zero-age main-sequence (ZAMS) mass of $M_{\rm ZAMS}\sim10-20$ \msun~\citep{jenson23,kilpatrick23, neustadt24,ransome24sn23ixf,vandyk24sn23ixf}. The ZAMS mass of the progenitor estimated using time variability of the IR emission is $20\pm4$ \msun, which if correct would make the progenitor of SN\,2023ixf one of the most massive progenitors of a Type II SN detected to date \citep{soraisam23}. Based on light curve modeling with the pre-explosion pulsation of the progenitor taken into account, \citet{hsu24} suggest the initial progenitor mass to be $M>17$ \msun.  On the other hand, \citet{bersten24} estimate a ZAMS mass of 12 \msun~with an explosion energy of $1.2\times10^{51}$ erg and a $^{56}$Ni mass of 0.05 \msun~based on comparison of the observations with their light curve models. 

The morphological light curve parameters measured by \citet{bersten24} using bolometric measurements show that SN\,2023ixf had a steeper decline during the plateau phase than the average SN II, putting it into the sub-class Type IIL. Previous studies show that morphological light curve parameters and spectroscopic properties of Type IIP and Type IIL form a continuous population, with a more general term Type II frequently used to refer to SNe falling under the two sub-classes \citep{anderson14, valenti16, dejaeger19}. \new{The morphological difference is explained either by SNe IIL having less massive H-rich envelopes after more vigorous mass-loss \citep{moriya16,Hiramatsu21} or by them having larger amounts of circumstellar material (CSM) around their progenitors at the time of explosion \citep{Morozova17}. While forming a continuous population, SNe IIL may originate from more massive progenitors than SNe IIP. \citet{chugai21} finds that SNe IIL have higher \nifs\ and confined circumstellar shell masses than SNe IIP, implying higher progenitor masses.}

The early-time (“flash”) spectroscopy of SN\,2023ixf revealed emission lines of H\one, He\one/\two, C\four, and N\three/\four/\five~ with narrow cores and broad, symmetric wings arising from the photo-ionization of dense, close-in CSM prior to shock breakout \citep{bostroem23, hiramatsu23, jacobson-Galan23, smith23ixf, teja23sn23ixf, yamanaka23, zimmerman24sn23ixf}. 
The early light curves and spectra indicate that the CSM was dense, produced by a mass loss rate of $\dot{M}$ = $10^{-3}-10^{-2}$~\msun~yr$^{-1}$, and concentrated near the stellar surface \citep[$R_{\rm CSM} \lesssim 10^{14}$ cm,][]{bostroem23,martinez24}. The mass loss rate estimated from the early SN observation or the pre-explosion imaging as previously mentioned, possibly hints at a complex mass-loss history prior to the explosion. The hard X-ray emission implies a number density of $n_{\rm CSM}$ = 4 $\times$ 10$^8$ cm$^{-3}$ at $r <$ 10$^{15}$ cm, corresponding to an enhanced progenitor mass-loss rate of 3 $\times$ 10$^{-4}$ \msun~yr$^{-1}$ for an assumed wind velocity of $v_w$ = 50 \kms\, which is much lower than the previously mentioned estimates based on the early light curves and spectra \citep{grefenstette23}.

Core-collapse supernovae (CCSNe) produce dust in their cooling ejecta. The time between the progenitor formation and the dust production is a few million years. This is substantially shorter than the time for dust to be produced by asymptotic giant branch (AGB) stars. AGB stars are the main dust producers in the present universe, but they cannot explain the amount of dust found in galaxies at $z\sim6-8$~($M_d\sim10^{6-8}$ $M_{\odot}$). This indicates CCSNe may be responsible for a large fraction of the dust in the early universe \citep{morgan03,dwek07,michalowski10,gall11,lesniewka19}. 
Large masses ($0.01-1$~\msun) of newly-formed dust have been found in nearby SN remnants, including Cas A, Crab Nebula, SN\,1987A, and G54.1+0.3 \citep{delooze17,delooze19,priestley20, matsuura15, temim17,rho18}, while much smaller amounts of dust ($10^{-6}-10^{-3}$ $M_\odot$) have been found in young CCSNe after only a few tens to hundreds of days after the explosion. Examples of the latter include SN\,2004dj \citep{meikle11,szalai11}, SN\,2010jl \citep{gall14,Fransson14}, SN\,2017eaw \citep{tinyanont19}, SN\,2020oi \citep{rho21}, and SN\,2021krf \citep{ravi23}. 

The formation of coolant molecules, including carbon monoxide (CO), precedes dust formation. CO is the first molecule to form in cooling ejecta because it has the highest dissociation energy of any molecule. Newly formed CO, which radiates efficiently in the infrared, cools the ejecta sufficiently for the dust particles to start to form. Near-infrared (NIR) emission from freshly formed CO and dust particles has been observed and analyzed in various types of CCSNe (e.g., Type II SN\,2017eaw, \citealt{rho18sn}; Type II-pec SN\,1987A, \citealt{spyromilio88,liu92,Wooden93}; Type Ic SN\,2020oi, \citealt{rho21}). Similarly, \citet{kotak09} obtained spectra of Type IIP SN\,2004et at near- and mid-infrared wavelengths, analyzing the evolution of the SiO molecules and the dust particles. By acquiring more examples of CO and dust emission in CCSNe, one can begin to look for relationships between progenitor type and amounts of CO and dust formation. 

In this paper, we present and describe 16 NIR spectra of SN\,2023ixf obtained with instruments at four telescopes, between 9 and 307 d after the explosion. In Section \ref{sec:obs}, we summarize the observations and data reduction procedures for each instrument. In Section \ref{sec:results}, we discuss intermediate-width (IW; $\sim$1000~\kms) line profiles found in the early epochs that may indicate the interaction between the expanding ejecta and the surrounding CSM. We also discuss the CO first overtone emission and the rising dust continuum found in the nebular phase spectra and estimate the physical properties of the gas containing the newly-formed CO molecules using an emission model assuming the CO is in a local thermodynamic equilibrium (LTE). We compare the CO and dust formation in SN\,2023ixf with that of other CCSNe reported in the literature, with a focus on SN\,1987A. The availability of the 16 NIR spectra and their related findings is critical to understanding JWST observations of SN\,2023ixf, which obtained spectra at only three epochs \citep[33, 252, and 373~d;][]{derkacyj25sn23ixf,medlerk25sn23ixf}.

%Figure1
\begin{figure*}
\centering
\includegraphics[width=17cm,height=12truecm]{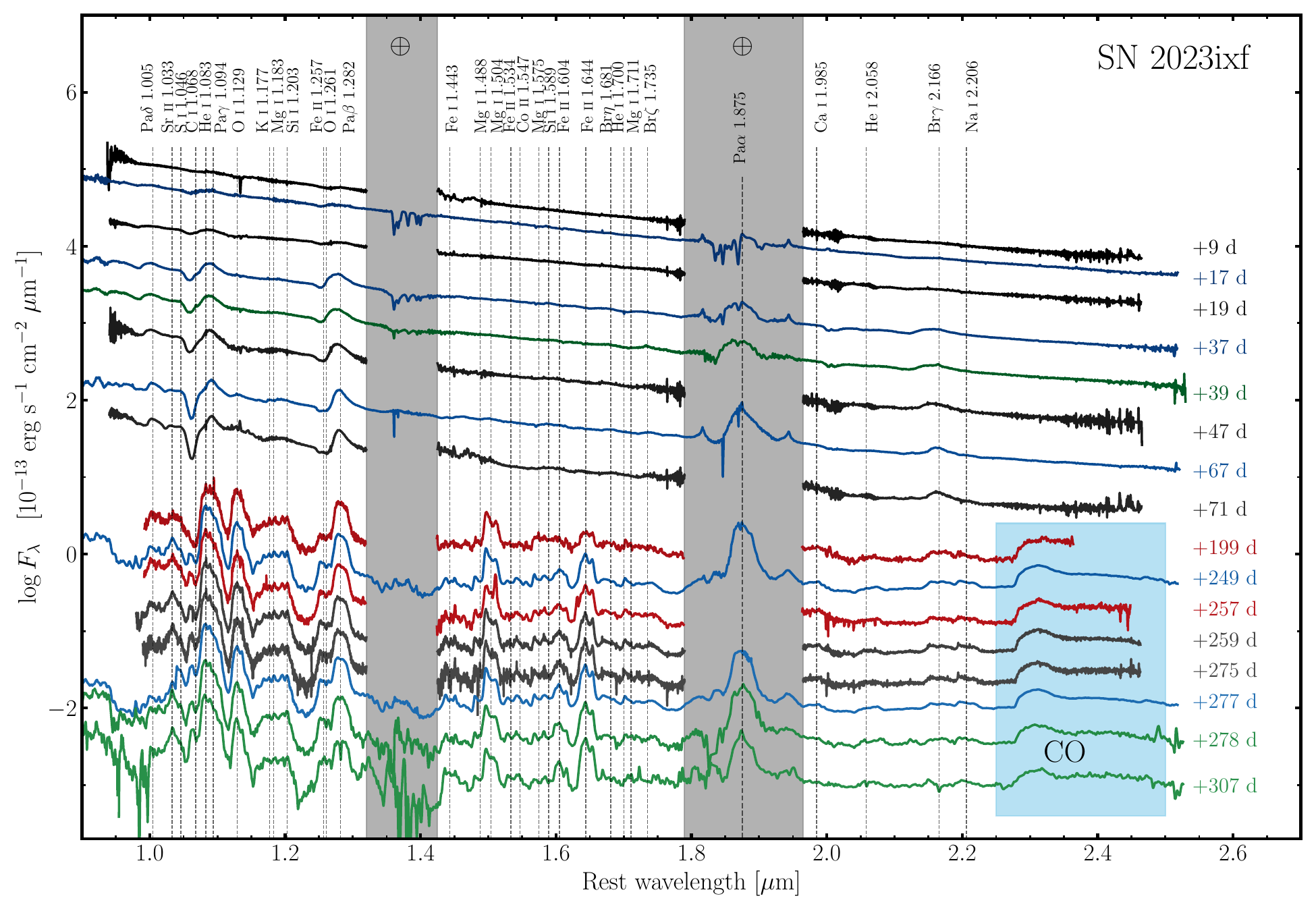}
\caption{Gemini/GNIRS (17, 37, 67, 249, and 277~d; blue), Keck/NIRES (9, 19, 47, 71, 259, and 275~d; black), IRTF/SpeX (39, 278, and 307~d; green), and MMT/MMIRS (199 and 257~d; red) spectra of SN\,2023ixf, in time order (top to bottom). Each spectrum is labeled with the days elapsed after the explosion. The wavelengths of known strong lines \citep[e.g.,][]{meikle89} are marked with dotted lines. Regions with significant noise due to strong telluric absorption are shaded in grey. The wavelength region of the first overtone CO emission is shaded in light blue. We also include Paschen $\alpha$ (1.875~\mic), which falls in a region of poor telluric transmission, marked by a black dotted line. The Brackett $\delta$ and $\epsilon$ lines, which also fall in the region, are artifacts due to incomplete removal of the same lines in the standard stars, despite appearing to be present in some of the spectra.
}
\label{nirspec}
\end{figure*}

\section{Observations} \label{sec:obs}

\subsection{Near-infrared Spectroscopy}
{\bf Gemini GNIRS observations:} We obtained NIR spectroscopy of SN\,2023ixf in five epochs spanning $17-277$~d after the explosion with the Gemini Near-Infrared Spectrograph (GNIRS) on the 8.1 m Frederick C. Gillett Gemini North telescope for programs GN-2023A-DD-105, GN-2023A-Q-218, GN-2023B-Q-225, and GN-2024A-DD-101 where GN-2023A-Q-218 and GN-2023B-Q-225 were from Gemini Korean time. We will compare these data with spectra of SN2017eaw obtained for GN-2017A-DD8 and GN-2017B-DD5, which were previously reported by \citet{rho18sn} but newly analyzed in this work.
For all of these observations of both SNe, GNIRS was configured in its cross-dispersed mode; the 32~line~mm$^{-1}$ grating and a 0.45 arcsec-wide slit (oriented at the mean parallactic angle during each observation), which provided a resolving power of $R\sim1200$ covering $0.84-2.52$ \mic. We utilized the GNIRS cross-dispersed reduction pipeline package \texttt{XDGNIRS} \citep{xdgnirs} for reducing the spectra. \cite{rho21} and \cite{ravi23} describe the data reduction steps in detail.

{\bf Keck NIRES observations: } We obtained six NIR spectra of SN\,2023ixf with the Near-Infrared Echellette Spectrometer (NIRES) on the Keck II telescope, as part of the ongoing Keck Infrared Transient Survey (KITS; \citealp{Tinyanont2024}). 
NIRES, a member of the TripleSpec family \citep{wilson2004}, provides an $R\sim2700$ spectrum with its fixed 0\farcs55 slit, over the wavelength range $0.9-2.45$ \mic.
The data were reduced using the \texttt{pypeit} general spectral reduction pipeline \citep{prochaska2020,pypeit2020}.
Further details of KITS observation and data reduction strategy can be found in \citet{Tinyanont2024}.

{\bf IRTF spectra:} We observed SN\,2023ixf three times with the short cross-dispersion (SXD) mode of the SpeX spectrograph \citep{rayner03} on the NASA InfraRed Telescope Facility (IRTF). Our use of the 0\farcs8-wide slit resulted in a resolving power of $R\sim750$. The SN was observed in an ABBA dithering pattern, as was an A0V star close in airmass. The associated flat-field and comparison arc-lamp observations were taken immediately after the astronomical observations. 
%\new{For the IRTF spectrum on 307~d, flux calibration using a standard star spectrum taken on the same night resulted in the overall flux being significantly underestimated. We reduced the spectrum using the same standard star observed on 278~d (with the same IRTF setup; See Table 1) and found the spectral shape remained unchanged. We believe the light from the standard star was not fully captured on 307~d, and we adjusted the fluxes using the new calibrator.}
%\new{For the 307~d spectrum, spectrum reduction using a standard star spectrum taken on the same day resulted in the flux being significantly underestimated. We reduced the spectrum using standard stars observed on different days and found the spectral shape remained unchanged. We believe the light from the standard star was not fully captured on 307~d, and adjusted the fluxes using the new calibrator.}
The data were reduced using Spextool \citep{cushing04}; the reduction steps are described by \cite{rho21}.

{\bf MMIRS spectra:} We obtained two sets of \textit{zJ} and \textit{HK} (taken using the high-throughput HK3 filter) spectra at 199 and 257~d post-explosion using the MMT and Magellan Infrared Spectrograph \citep[MMIRS;][]{mmirs} on the 6.5~m MMT located on Mt.\ Hopkins in Arizona. We also obtained one K3000 ($1.90-2.45$ \mic) spectrum at 257~d. All MMIRS spectra were taken using a 1.0$''$-wide longslit. For data reduction, we used the MMIRS pipeline \citep{mmirspipe}. Telluric and absolute flux corrections were done according to the method described in \citet{vacca03} with XTELLCOR \citep[a part of Spextool package; ][]{cushing04}, using a standard A0V star observed at similar airmass and time. 

In total, 16 NIR spectra of SN\,2023ixf were obtained using four different telescopes. The spectra are shown in Figure \ref{nirspec}. The information for the NIR observations of SN\,2023ixf is summarized in Table \ref{Tobs}.
We adopt a redshift of $z=0.000804$ for SN\,2023ixf from \citet{hosseinzadeh23} for M101and an extinction value of 
$E(B-V)_{\rm host}$ = 0.031 ± 0.006 mag \citep{hosseinzadeh23, smith23ixf}, which has a negligible effect on the NIR spectra.

\subsection{Optical Spectroscopy}
LCO optical spectra from 6 to 440~d after the explosion were taken with the FLOYDS spectrographs mounted on the 2m Faulkes Telescope North (FTN)  at Haleakala (USA) through the Global Supernova Project. The first three spectra from 6, 7, and 9 d were previously published in \citet{bostroem23}. All spectra were obtained through a 2$''$-wide slit oriented at the parallactic angle in order to minimize light loss \citep{filippenko82}. One-dimensional spectra were extracted, reduced, and calibrated following standard procedures using the FLOYDS pipeline\footnote{\url{https://github.com/svalenti/FLOYDS_pipeline}} \citep{valenti14}. The oscillations in the FLOYDS spectra that appear above 7500 \AA, are due to fringing.
We focus on hydrogen recombination lines that show absorption features as shown in Figure \ref{HVplot}. The entire set of optical spectra and their analysis will be presented in another paper (Brian Hsu, in prep).

%Table1
\begin{table*}
\scriptsize
\caption{Summary of IR Observations}\label{Tobs}
\begin{center}
\begin{tabular}{cccccccccc}
\hline \hline
No. & Date   &  Telescope       &  MJD (phase)$^a$    &  CO mass & CO T & CO FWHM & $J$, $H$, $K^b$ & T$_d$$^g$ & M$_d$$^h$\\
 & (yyyymmdd) &     &  & ($10^{-4}M_{\odot}$) & (K) & (km s$^{-1}$) & (mag) & (K) & ($10^{-5}M_{\odot}$)\\
 \hline \hline
 & {\bf SN\,2023ixf} &&&\\
1 & 20230528  & Keck       & 60092 (9)   & ... & ... & ... & 11.68, 11.60, 11.49 & ...& ...\\
%Intermediate + Broad P Cygni Pa beta& \\
2 & 20230605  & Gemini & 60100 (17)   & ... & ... & ...& 10.63, 10.52, 10.33 & ...& ...\\
3 & 20230607  & Keck       & 60102 (19)  & ... & ... & ...& 11.04, 10.78, 10.50 & ...& ...\\
%strong H lines & \\
4 & 20230625  & Gemini & 60120 (37)   &... & ... & ...& 11.07, 10.92, 10.66 &...& ...\\
%H+He + narrow lines & \\
5 & 20230627  & IRTF        & 60122 (39)   & ... & ... & ...& 10.94, 10.84, 10.58 &...& ... \\
6 & 20230705  & Keck       & 60130 (47)   & ... & ... & ...& 11.40, 11.35, 11.12 & ...& ...\\
7 & 20230725  & Gemini & 60150 (67)  &  ... & ... & ...& 11.25, 11.09, 10.82 &...& ... \\
%narrow lines at 1.25 and 1.92um & \\
8 & 20230729  & Keck       & 60154 (71)   & ... & ... & ...& 11.71, 11.73, 11.53 & ...& ...\\
\hline
  &Model A$^c$& &\\
9 & 20231204  & MMT           & 60282 (199)   & (2.3)$^d$ & (2470)$^d$ & (3370)$^d$ & 14.31, 14.46, 13.59 & ...& ...  \\
10& 20240123  & Gemini & 60332 (249)  & $2.66\pm0.28$ (1.8-4.6)$^e$ & $1775\pm47$$^e$ & $3492\pm231$ &  15.31, 14.97, 13.86 & & \\
11& 20230131  & MMT          & 60340 (257)   & (3.1)$^f$ & (1970)$^f$ & (4050)$^f$ & 14.77, 15.16, 13.49 & ...& ...\\ 
12& 20240202  & Keck       & 60342 (259)    & $1.38\pm0.24$ (0.9-2.3)$^e$ & $1741\pm76$$^e$ & $3587\pm292$ & 16.03, 15.74, 14.69 & ...& ...\\
13& 20240218  & Keck       & 60358 (275)    & $0.74\pm0.10$ (0.5-1.1)$^e$ & $1996\pm77$$^e$  & $3162\pm263$ & 16.12, 15.76, 14.73 & ...& ...\\
14& 20240220  & Gemini     & 60360 (277)    & $1.78\pm0.24$ (0.9-2.5)$^e$ & $1669\pm53$$^e$  & $3248\pm264$ & 16.01, 15.53, 14.58 & ...& ... \\
15& 20240221  & IRTF       & 60361 (278)   & $0.93\pm0.16$ (0.7-1.3)$^e$ & $2098\pm102$$^e$  & $3016\pm412$ & 15.76, 15.25, 14.21 &...& ...\\
%16 & 20240322 & IRTF/SpeX        & 60390 (307)    & $0.05\pm0.01$$^e$ & $2143\pm96$$^e$ & $3581\pm436$ & 18.56, 18.06, 17.12 & ...& ...\\
%16 & 20240322 & IRTF        & 60390 (307)    & $0.15\pm0.02$ (0.1-0.2)$^e$ & $3061\pm123$$^e$ & $3837\pm563$ & 16.54, 16.00, 15.00 & ...& ...\\
16 & 20240322 & IRTF        & 60390 (307)    & $0.39\pm0.06$ (0.3-0.6)$^e$ & $2143\pm96$$^e$ & $3581\pm436$ & 16.44, 15.93, 14.99 & ...& ...\\
\hline &Model B$^c$&&&\\
9 & 20231204  & MMT          & 60282 (199)    & (2.3)$^d$ & (2480)$^d$ & (3370)$^d$ & 14.31, 14.46, 13.59 & (1044)$^f$& (2.25)$^f$\\%       
10 & 20240123  & Gemini & 60332 (249)  & $2.01\pm0.19$ (1.4-3.1)$^e$ & $1922\pm47$$^e$  & $3319\pm221$ &  15.31, 14.97, 13.86 & 915$^g$ & 2.25$^h$ \\
11 & 20230131  & MMT          & 60340 (257) & (2.9)$^f$ & (2020)$^f$ & (4020)$^f$ & 14.77, 15.16, 13.49 & (1014)$^f$ & (1.62)$^f$\\ 
12 & 20240202  & Keck       & 60342 (259)    & $1.24\pm0.21$ (0.8-2.0)$^e$ & $1798\pm76$$^e$ & $3533\pm287$ & 16.03, 15.74, 14.69 & 968$^g$& 0.74$^h$ \\
13 & 20240218  & Keck       & 60358 (275)    & $0.67\pm0.09$ (0.5-1.0)$^e$ & $2065\pm78$$^e$  & $3152\pm258$ & 16.12, 15.76, 14.73 & 955$^g$ & 0.80$^h$\\
14 & 20240220  & Gemini     & 60360 (277)    & $1.30\pm0.15$ (0.9-2.1)$^e$ & $1820\pm52$$^e$  & $3092\pm246$ & 16.01, 15.53, 14.58 & 931$^g$ & 1.41$^h$\\
15& 20240221  & IRTF        & 60361 (278)   & $0.75\pm0.12$ (0.6-1.0)$^e$ & $2247\pm104$$^e$  & $3010\pm391$ & 15.76, 15.25, 14.21 & 1036$^g$ & 0.77$^h$ \\
%16& 20240322  & IRTF/SpeX        & 60390 (307)    & $0.20\pm0.03 [SH. ERRORS LOOK STILL TOO SMALL CONSIDERING THE SPECTRA ARE VERY NOISTY) $($^e$ & $2802\pm130$$^e$  & $3813\pm514$ & 16.54, 16.00, 15.00 & 1059$^g$ & 0.33$^h$ \\
%16& 20240322  & IRTF        & 60390 (307)    & $0.20\pm0.03$ (0.1-0.3)$^e$ & $2802\pm130$$^e$  & $3813\pm514$ & 16.54, 16.00, 15.00 & 1059$^g$ & 0.33$^h$ \\
16& 20240322  & IRTF        & 60390 (307)    & $0.31\pm0.04$ (0.2-0.4)$^e$ & $2305\pm97$$^e$  & $3520\pm423$ & 16.44, 15.93, 14.99 & 1026$^g$ & 0.40$^h$ \\
\hline
& {\bf SN\,2017eaw}$^i$&&&\\
 & 20170915 &Gemini & 58011 (124)  & 1.58$^{+0.13}_{-0.07}$ (0.6--1.6) & 2750$^{+200}_{-220}$ (3000$\pm$200)$^j$ & 2360$^{+210}_{-270}$ (2800$\pm$200)$^j$ &... &...& ...\\
 & 20171001  &Gemini & 58027 (140)  & 1.61$^{+0.19}_{-0.02}$ (1.0--1.9) & 3060 $^{+280}_{-450}$ (3300$\pm$200)
 & 2724$^{+265}_{-515}$ (2850$\pm$200) & ... &...& ...\\
 & 20171030  &Gemini & 58056 (169)  & 1.95$^{+0.09}_{-0.04}$ (1.6--2.2) &  2915$^{+140}_{-170}$ (3000$\pm$200) & 2750$^{+135}_{-185}$ (2850$\pm$200) &...  &... & ... \\
 & 20171205  &Gemini & 58092 (205)  & 2.21$^{+0.16}_{-0.11}$ (1.9--2.2) & 2584$^{+110}_{-130}$ (2700$\pm$200) & 2735$^{+115}_{-145}$ (2750$\pm$200) &...& ...&...  \\
\hline
\end{tabular}
\end{center}
\renewcommand{\baselinestretch}{0.8}
\footnotesize{$^a$ t$_0$ = 60083 MJD is taken to be the explosion date of SN\,2023ixf.}\\
\footnotesize{$^b$ $J, H, K$ band magnitudes were measured using the NIR spectra.}\\
\footnotesize{$^c$ CO models were fit adopting a flat continuum (the elevated “pseudo-continuum”) at 2.0–2.3 \mic\ for Model A and a modified black body continuum with carbon dust for Model B.}\\
\footnotesize{$^d$ This spectrum did not cover enough of the CO band emission to allow errors to be estimated.}\\
\footnotesize{$^e$ The errors provided are statistical. In addition to the standard statistical errors provided in the CO mass and temperature columns, the 10-20\% uncertainty in the flux value and the telluric correction may result in uncertainties of 100-200 K in temperatures in the fitting process. \new{The CO mass ranges (in parentheses) are estimated, accounting for the error in ${\rm T}_{\rm CO}$ of 200 K.}}\\ 
\footnotesize{$^f$ This spectrum was obtained using the K3000 filter (which does not cover the full CO feature) and may be subject to systematic errors due to the novel setup. Additionally, errors in the dust mass could not be estimated.}\\
\footnotesize{$^{g,h}$ Statistical errors of the dust temperatures and mass are $\sim$35~K and $\sim~0.15\times10^{-5}$ M$_{\odot}$, respectively. \new{The systematic errors in the temperatures are $\sim$100~K which results in the uncertainties of the dust mass $\sim$0.8$\times10^{-5}$ M$_{\odot}$ (a factor of $\sim$3 and $\sim$5 higher, respectively), due to uncertainties in defining the continuum.}}\\
%The difference in the dust temperatures may not be significant due to the presence of lower-temperature components.}\\ 
\footnotesize{$^i$ The uncertainties in the model fitting include statistical and systematic errors resulting from different choices of the continuum levels.}\\
\footnotesize{$^j$ The values in parentheses are from \citet{rho18sn} for comparison.}
\end{table*}

%Table2
\begin{table*}%[!hb]
\scriptsize
\caption{Observed Line Properties of Narrow Features and Associated Broad Lines}
\begin{tabular}{cccccc|ccccccl}
\hline
Line ($\lambda_0$) & Phase & \multicolumn{4}{c}{Intermediate-Width (IW)$^a$ or Narrow$^b$ line} & \multicolumn{4}{c}{Broad or P Cygni line}\\
 &    & \multicolumn{2}{c}{Emission line} &  \multicolumn{2}{c}{Absorption line} &\multicolumn{2}{c}{} &\multicolumn{2}{c}{FWHM}\\
   &   &  $V_{\rm shift}$ & FWHM & $V_{\rm shift}$ & FWHM & $V_{\rm shift}$ & $\Delta V$$^c$ & Absorption & Emission\\
   (\mic) & (day) & (\kms)\  &  (\kms)\  & (\kms)\  &(\kms)\  & (\kms) & (\kms) & (\kms) & (\kms)\\
\hline \hline
He\one~(1.083) & 9  & 0 & 670$^a$ & -530 & 1000$^a$ & $-6450$ & 7300 & 5250 & 6010 \\
(Pa\,$\gamma$ 1.094) & 9 & (0) & (980)$^a$ & ... & ... & ... & ... & ... & ...\\
He\one~(1.083) & 17 &  ... & ... & $-370$ & 600$^a$ & $-6200$ & 7900 & ... & ...\\
 & 19 & 0 & 500$^a$ & ... &... &$-6700$ & 8300 & ... & ...\\
(C\one\ 1.0695) & (39) & -400 & 400$^a$ &... & ... & ... & ...& ... & ...\\
He\one~(1.083) & 47 & 0 & 100$^b$ & $-50$ & 100$^b$& $-6900$ & 8300 & ... & ...\\
 & 67 & ... & ... & $-150$ & 200$^b$ & $-6000$ & 8200 & ... & ...\\
 & 71  &0 & 50$^b$ &$-100$ & 100$^b$ & $-5800$ & 8100 & ... & ...\\
\hline
O\one~(1.129) & 17 &-150 & 500$^a$ & ... & ... & ... & ...& ... & ...\\
\hline
Pa\,$\beta$ (1.282) & 9 & $-80$ & 720$^a$ & ... & ... & $-5170$ & 6300 & 4790 & 5150 \\
\hline
Br\,$\gamma$ (2.166) & 17 & 200 & 370$^a$ & ... & ... &$-7000$ & 7800 & ... & ...\\
 & 39 & $-270$ & 600$^a$ &...&... & $-7550$ & 6400 & ... & ...\\
\hline 
\end{tabular}\\
\centering
\footnotesize{$^a$ Lines at $t=9-39$~d with FWHMs of $\sim$200--1000~\kms.}\\
\footnotesize{$^b$ Lines at $t=47-71$~d with FWHMs of $\lesssim$200~\kms.} \\
\footnotesize{$^c$ Width of velocity interval between minimum and maximum of P Cygni profile.   }
\label{Tlinefluxes}
\end{table*}

\section{Results and Discussion} \label{sec:results}
\subsection{Near-infrared Spectra of SN\,2023ixf}
Figure \ref{nirspec} shows the sixteen sets of 0.8--2.5~$\mu$m spectra of SN\,2023ixf between days 9 to 307 since the explosion. 
These data form one of the most extensive sets of near-infrared spectra of a Type II supernova obtained to date. The only other comparable NIR coverage of a Type II SN is that of SN\,2017eaw \citep{rho18sn}. The near-infrared spectra set from SN\,2023ixf covers 9 -- 71~d and 199 -- 307~d, while that of SN\,2017eaw covers $t<$205~d. We are making the sixteen SN\,2023ixf NIR spectra available to the astronomy community via VizieR.

The first eight spectra, obtained during the photospheric phase, are dominated by hydrogen recombination line emission, a [S~II] absorption at 1.046 \mic, and He absorption features (weaker than those of SN\,2017eaw) all superimposed on a continuum that decreases monotonically with increasing wavelength. 
In the spectra between 199 and 307~d, some of the line emission is from gas that is nebular in density. The atomic lines become much more prominent in the nebular phase.
In addition, CO overtone band emission at 2.3--2.5~\mic~is present in these latter spectra, together with a flattening of the continuum at $\lambda>1.5$ \mic, not seen in the earlier spectra, which we interpret as emission from newly formed hot dust.
  
\subsection{Line Profiles and CSM properties}

The earliest few optical spectra of SN\,2023ixf exhibited narrow (FWHM $\lesssim200$ \kms) hydrogen emission lines and narrow flash-ionized lines (e.g., He\two, C\four, N\four) that disappeared within one week after the explosion \citep{smith23ixf,bostroem23,jacobson-Galan23}. Intermediate-width (IW; 200~\kms $\lesssim$ FWHM $\lesssim$ 1000~\kms) P Cygni \Ha\ profiles were observed in the days following the disappearance of narrow lines before they themselves disappeared by 18~d \citep{smith23ixf}. The narrow lines in the early spectra are often attributed to dense CSM concentrated near the progenitor star, and the disappearance of the narrow lines is attributed to the ejecta engulfing the dense CSM \citep{terreran22_20pni,andrews24_22jox,dastiar24_19nyk}. The IW lines imply ejection velocities much faster than typical RSG wind velocities. \citet{smith23ixf} suggest that the IW features found in the optical are emitted from the cool, dense shell (CDS) formed at the ejecta-wind interface, similar to what has been observed in Type IIn SNe and Type IIP SNe interacting with CSM \citep{chugai04,andrews10,smith10,andrews11,andrews17}.
However, the same features may instead originate, as \citet{bostroem23} suggested, from unshocked and radiatively accelerated CSM in front of the shock if the surrounding CSM is dense and sufficiently confined \citep{chugai02,tsuna23}.

%Figure2
\begin{figure*}%[!h] %[width=0.48\textwidth
\begin{tabular}{cc}
\includegraphics[width=0.47\textwidth, height=11.0truecm]{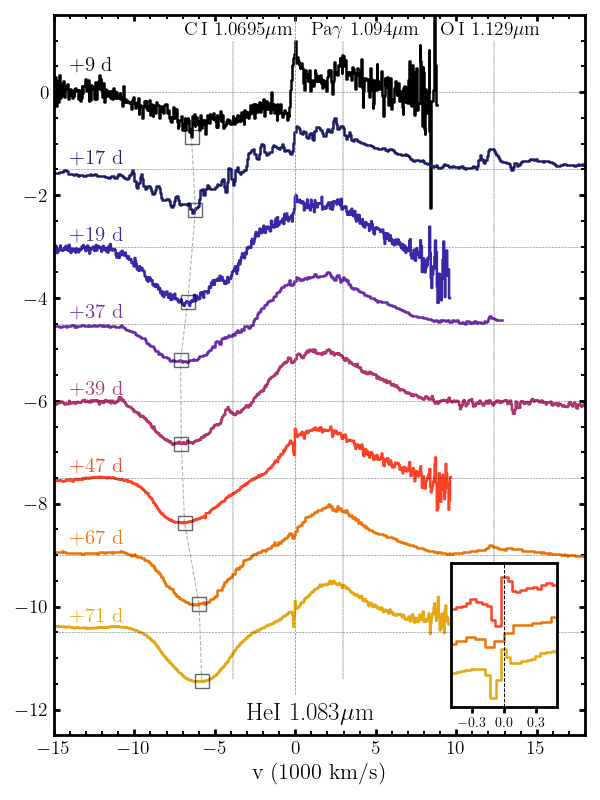}&
\includegraphics[width=0.47\textwidth, height=11.0truecm]{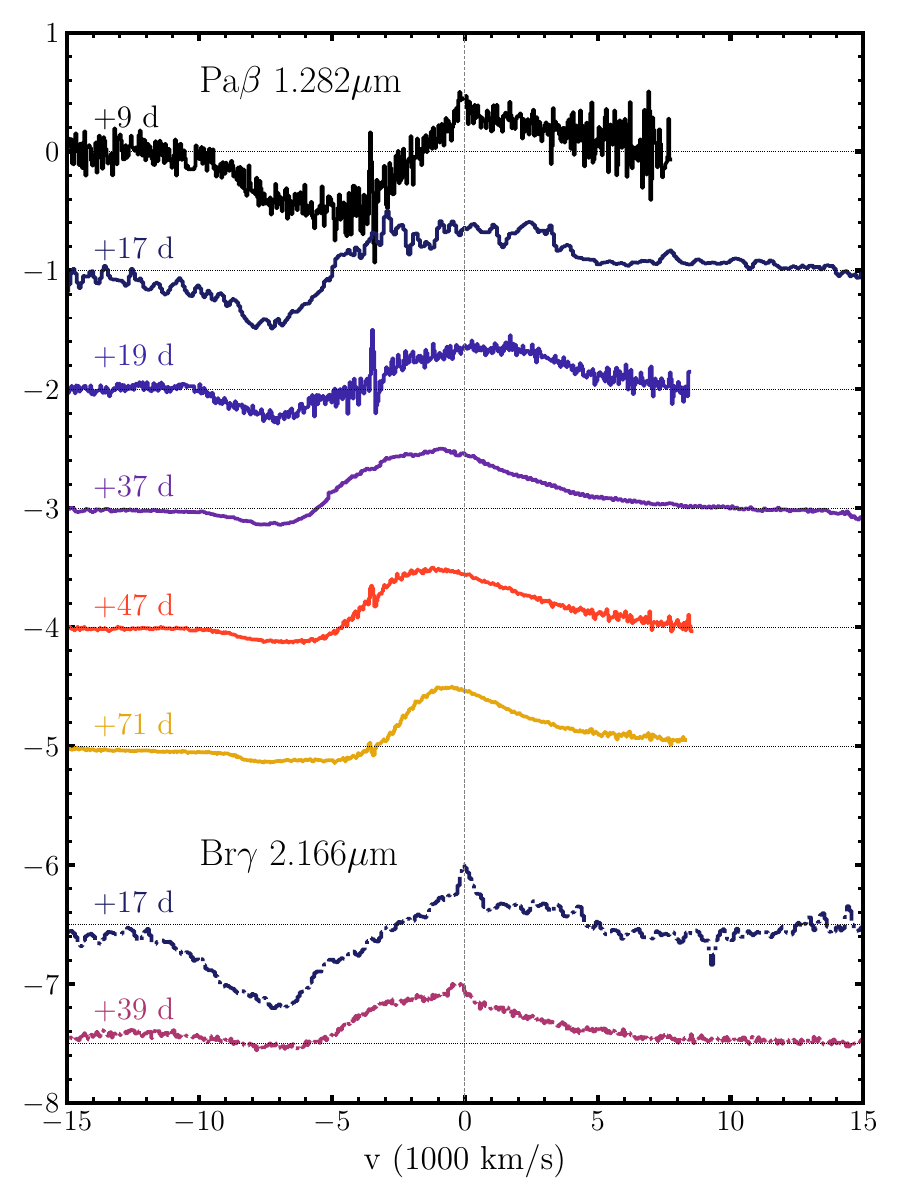}\\
(a) & (b) \\
\end{tabular}
\caption{\textit{(a)} Velocity profiles of He\one~1.083 \mic~from 9 to 71~d. Lines identified from intermediate-width (IW) features (FWHM$\sim$1000 \kms) are marked with black dotted vertical lines. The inset at the bottom right corner shows narrow (FWHM$\sim$100 \kms) features at 47, 67, and 71~d, near the zero velocity of He\,\one~1.083 \mic. \textit{(b)} Velocity profiles of Pa\,$\beta$ 1.282 \mic~and Br\,$\gamma$ 2.166 \mic~from selected epochs. The Pa\,$\beta$ line shows an IW feature at 9~d, and the Br\,$\gamma$ line shows IW features at 17 and 39~d. 
See Table \ref{Tlinefluxes} for the FWHMs and the shifts from zero velocity of the IW and narrow lines under the columns labeled $`$Narrow line', and also show the velocity shifts of broad absorption, and the velocity interval between minimum and maximum of P Cygni profile.}
\label{vplotHeI}
\end{figure*}

\begin{figure*}[!h] %[width=0.48\textwidth
\begin{tabular}{cc}
\includegraphics[width=0.47\textwidth, height=11.0truecm]{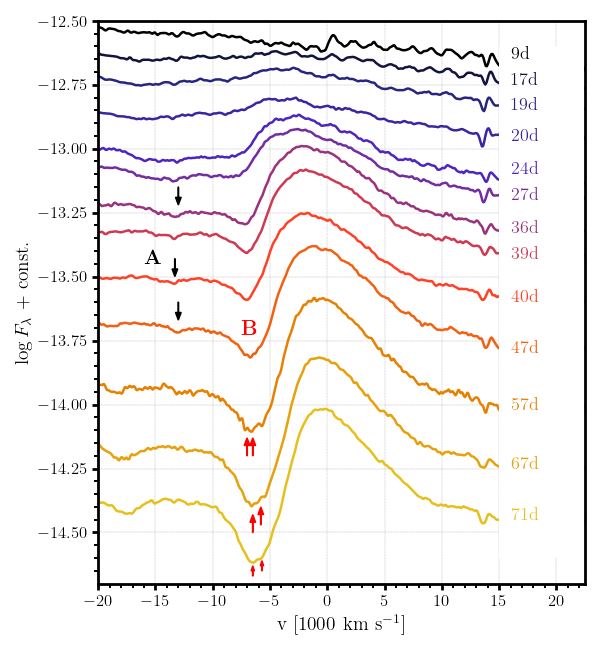}&
\includegraphics[width=0.47\textwidth, height=11.0truecm]{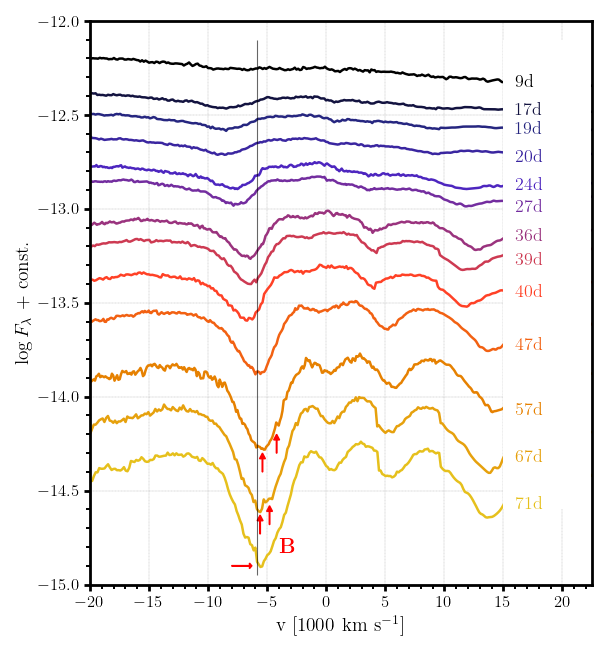}\\
(a) & (b) \\
\end{tabular}
\caption{
H\,$\alpha$ \textit{(a)} and H\,$\beta$ \textit{(b)} line profiles of SN\,2023ixf; Of particular interest are a “dip” ('B' marked in red) on the blue side of the profile with velocities of $\sim$-6000~\kms~at 57, 67, and 71~d, and a moderately strong absorption feature ('A' marked in black) at high velocity (-12,000 \kms) between 27 and 57~d \citep[see also][the HV in SN 1999em]{leonard22}. \new{These features (indicated by arrows and labeled 'A' and 'B') are called $`$Cahitos' \citep[see ][]{gutierrez17}. These feature are caused by interactions, geometry, and viewing angles between the reverse shock, the CSM and ejecta \citep{chugai04}.} The epochs are labeled on the right. The absorption-like features at 14,000 \kms\ in H\,$\alpha$ are due to telluric absorption.}
\label{HVplot}
\end{figure*}

Our set of NIR spectra, presented in Figure \ref{nirspec}, includes only two epochs (9 and 17~d) before 18~d when the optical spectrum was almost featureless. The optical spectra near 17~d show weak broad absorptions without noticeable broad emissions, but the NIR spectra show broad features, many of which contain both absorption and emission, starting at 9~d. We also find sporadic narrow features well into the photospheric phase ($t\gtrsim30$~d), which may indicate episodic interactions between wind clumps and CSM.

A list of narrow line features and the associated velocities is presented in Table \ref{Tlinefluxes}. The table includes the widths of the narrow features, the velocity shifts measured at the peak of the emission and the absorption features for both narrow and broad features, if present. Figure \ref{vplotHeI} contains velocity profiles of these narrow features, with the y-axis being the flux normalized to the continuum. Narrow He\one~features were present from 9~d to 71~d. The differences in FWHMs between the narrow features in the earlier epochs ($\sim$20 d) and in the later epochs ($39-71$~d) may indicate different origins of the features in these two time periods. \new{In the paragraphs following, we distinguish IW features from narrow features according to their FWHMs. The FWHMs of the IW features estimated from the earlier epochs ($500-1000$~\kms) are consistent with the $``$intermediate-width" features in the optical spectra at those times, whereas the FWHMs of the narrow features from the later epochs are smaller and are closer to those of narrow flash-ionized lines \citep{smith23ixf, bostroem23}.} 
It should be noted that the broad features in the region ($1.06-1.10$~\mic) are a blend of the He\one~1.083 \mic~and Pa\,$\gamma$ 1.094 \mic~lines with the relative contributions of the two changing with time, so interpreting the temporal evolution of $V_{\rm shift}$ or $\Delta V$ as the temporal evolution of the ejecta velocity is risky. 

\begin{figure*}
\begin{tabular}{cc}
\includegraphics[width=0.48\textwidth]{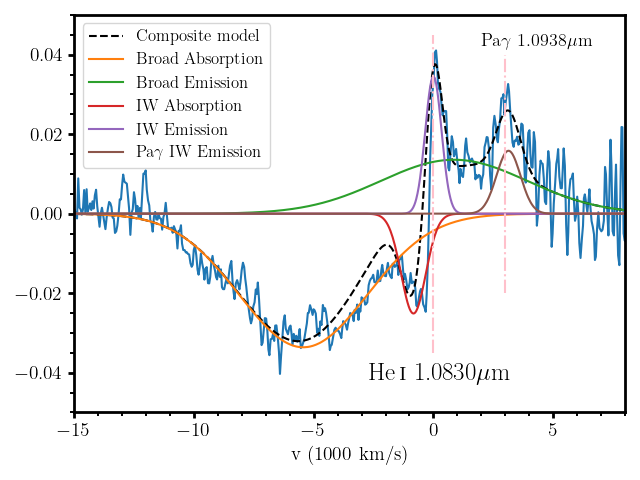}&
\includegraphics[width=0.48\textwidth]{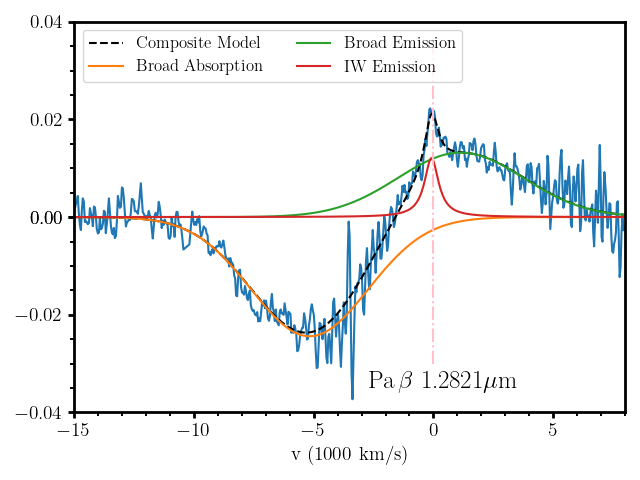} \\
(a) & (b)\\
\end{tabular}
\caption{Multi-component line fitting of He\one~1.083 \mic~(\textit{a}) and Pa\,$\beta$ 1.282~\mic\ (\textit{b}) at 9~d with with Gaussian and Lorentzian components. The He\one~1.083 \mic ~line profile (\textit{a}) was fitted with two broad Gaussian profiles with FWHMs of 5250 and 6010~\kms~for absorption and emission, along with two intermediate-width (IW) Gaussian profiles with FWHMs of 1000 and 670~\kms~for absorption and emission.
The Pa\,$\gamma$ 1.094 \mic~that overlaps the He\one\ 1.083 \mic~was simultaneously fitted with an intermediate-width Gaussian profile with FWHM of 980~\kms. The Pa\,$\beta$ 1.282~\mic\ line profile (\textit{b}) was fitted with two broad Gaussian profiles with FWHMs of 4790 and 5150~\kms~for absorption and emission, respectively, and an intermediate-width Lorentzian emission profile with FWHM of 720~\kms~(red).}
\label{multicomponentLinefitting}
\end{figure*}

We modeled the line profiles using multiple Gaussian or Lorentzian components in order to estimate the velocity shifts and the FWHM velocities presented in Table \ref{Tlinefluxes}. We used Gaussian profiles for broad (FWHM$\sim$5000 \kms) components, Lorentzian or Gaussian profiles for IW components.  Our choice between Lorentzian and Gaussian profiles depends on the shape of the lines. Figure \ref{multicomponentLinefitting} contains examples of multi-component fitting. In the right panel of Figure \ref{multicomponentLinefitting}, the Pa\,$\beta$ feature at $t=9$~d is fitted with three components: (1) a broad absorption with a Gaussian profile centered at a velocity shift of $\sim-5170$ \kms, (2) a broad Gaussian emission, and (3) an IW Lorentzian emission component centered at $\sim-80$ \kms, with a FWHM of 720~\kms. The fitting was performed sequentially, starting with the broad absorption component, followed by the broad emission component, and finally the IW component. In the case of an IW component from another line (e.g., Pa\,$\gamma$ and C\one) overlapping the major line feature, we performed the IW line fitting as the final step. This is shown in the left panel of Figure \ref{multicomponentLinefitting} where we fit the He\one~velocity plot with an overlapping Pa\,$\gamma$ IW emission shown at $t=9$~d with five Gaussian profile components: 1) broad absorption, 2) IW absorption, 3) broad emission, 4) IW emission, 5) IW Pa\,$\gamma$ emission.

P Cygni line profiles such as in H\,$\alpha$ lines indicate that the emitting region is very close to the photosphere, which causes a significant flux-deficit in the red, made greater by the rapidly declining density distribution \citep{dessart05spec}. This provides a clear explanation for the noticeable blue-shift of emission peaks of P Cygni profiles observed in early-time spectra of SNe II.
Line transfer effects in the ejecta above the photosphere produce P Cygni profiles whose shape and intensity depend upon the velocity and optical depth structures of the ejecta.
Analyses of the line profiles by combining multiple Gaussian and Lorentzian components instead of properly modeling synthetic spectra are, at best, phenomenological and should be viewed with caution. 
CSM interaction is another factor that further complicates analyses of P Cygni profiles. A cool dense shell formed by the reverse shock when the forward shock reaches the CSM may have produced the IW line emission and even IW absorption if a thick enough CSM existed near the progenitor surface before the explosion. The double P Cygni structure of He\one~1.083~\mic~at 9~d shown in the right panel of Figure \ref{multicomponentLinefitting} may imply a complex ejecta structure during the early ($\sim$10~d) phase of SN\,2023ixf.

At 47 and 71~d, weak and narrow P Cygni profiles were present in the He\one~1.083 \mic~line near $V\sim0$. Their FWHMs ($\sim$100 \kms) are lower than those of IW features from the earlier epochs but still imply ejection velocities faster than the typical velocity of a RSG wind. They may be due to the radiative acceleration of unshocked CSM or a faster wind ejected from the progenitor star sometime before the explosion \citep{smith23ixf}. Assuming an SN ejecta velocity of $\sim$10,000 \kms, and a pre-explosion wind velocity of 100 \kms, it can be inferred that the wind interacting with the ejecta during this epoch was ejected from the progenitor roughly 20 years prior to the explosion. For a 15 \msun~progenitor, this corresponds to the carbon shell burning stage \citep{fuller17}.

We also find IW features in lines other than He\one. At 9~d, the IW Pa\,$\gamma$ line overlaps the red wing of the broad He\one~emission.
At 39~d, an IW emission peak centered at 1.068~\mic~rises from the broad absorption trough of He\one. This line is near the C\one~multiplet at 1.0695~\mic, previously found in SN\,1987A at day 18 by \citet{meikle89} based on the prediction of \citet{branch87LTE87A}. The emission feature has disappeared by day 112 in SN\,1987A when He\one~1.083~\mic~line dominates at its wavelength. The C\one~line would be expected to reach maximum strength at a temperature of $\sim$4500~K and weaken as the ejecta cools \citep{meikle89}. The appearance of the C\one~line at 39~d for SN\,2023ixf is later than that of SN\,1987A, but the blackbody temperature of $\sim$5000~K found for SN\,2023ixf by \citet{singh24sn23ixfpre} at $\sim$39~d allows for the possible formation of this line. We find a weak emission bump and a broad excess at approximately the same wavelength at 19~d that may be related to this IW emission feature. Other IW features are present in the 9 to 17~d spectra. In Figure \ref{vplotHeI}b, the Pa\,$\beta$ 1.282 \mic~line at 9~d contains an IW feature on top of a broad P Cygni profile, which disappears by 17~d. Although a broad P Cygni profile strengthens at later epochs, no narrow or IW features associated with CSM interaction are present. Narrow features at $v\sim-4000$ \kms~($\lambda=1.267$ \mic) relative to Pa\,$\beta$ are artifacts of the reduction process due to strong telluric O$_{2}$ emission. At 17~d, an IW O\one~1.129 \mic~line not associated with a broader feature and a Br\,$\gamma$ profile with a weak, IW component on top of a broad P Cygni profile are present. The same complex Br\,$\gamma$ profile is still present at 39~d. The FWHM velocity and the blue-shift of IW Br\,$\gamma$ feature are comparable to those of the IW C\one~1.0695~\mic~line observed on the same day.

The IW features found before 20~d in NIR are similar to those found in the optical, except that the He\one~profiles also include IW absorption. The time of the disappearance of these NIR features is later than in the optical, with the Br\,$\gamma$ and a possible C\one~feature with FWHM~$\sim400-600$~\kms~being present in the 39~d spectrum, whereas the IW optical features disappear by 18~d \citep{smith23ixf}. We also find narrow absorption features of He\one~whose FWHM is $100-200$~\kms~with two of them accompanied by narrow emission features, together resembling P Cygni profiles.

%Figure3

\subsection{Optical high-velocity absorption features}
\begin{table}%[!htb]
\caption{Summary of Optical Spectroscopy of SN\,2023ixf using LCO}
\label{Topti}
\begin{center}
\begin{tabular}{llllc}
\hline \hline
No. & Date & MJD & phase (Day)$^a$& \\
\hline
1 &20230525 & 60089.33 & 6 & \\
2 &20230526 & 60090.25 & 7 & \\
3 &20230528 & 60092.28 & 9 & \\
4 &20230605 & 60100.43 &17 & \\
5 &20230615 & 60110.32 &27 & \\
6 &20230621 & 60116.36 &33 & \\
7 &20230627 & 60122.29 &39& \\
8 &20230705 & 60130.31 &47& \\
9 &20230715 & 60140.38 &57& \\
10 &20230729 & 60154.26 &71& \\
11 &20230815 & 60171.30 & 88& \\
12 &20230830 & 60186.23 & 103& \\
13 &20230909 & 60196.22 & 113 & \\
14 &20231203 & 60281.61 &198 & \\
15 &20231227 & 60305.60 &222& \\
16 &20240121 & 60330.49 & 247 & \\
17 &20240207 & 60347.44 &264& \\
18 &20240228 & 60368.49 & 285& \\
19 &20240319 & 60388.38 & 305& \\
20 &20240410 & 60410.57 & 327& \\
21 &20240525 & 60455.40 & 372 & \\
22 &20240611 & 60472.45 & 389& \\
23 &20240629 & 60490.38 & 407& \\
24 &20240801 & 60523.30 & 440& \\
\hline
\end{tabular}
\end{center}
\renewcommand{\baselinestretch}{0.8}
{\footnotesize $^{a}$ t$_0$ = 60083 MJD is the explosion date of SN\,2023ixf.}
\end{table}

The H\,$\alpha$ and H\,$\beta$ line profiles in SN\,2023ixf are shown in Figure \ref{HVplot}. High-velocity (HV) absorption features are present in many of the LCO optical spectra. 
The features (marked $`$A' and $`$B' in Figure \ref{HVplot}) were present on the blue sides of the P Cygni profile.
Similar HV features have been reported in SN\,2009N and a few other SNe, including SN\,1992ba \citep{takats14, gutierrez17, davis19}. 
These HV features are known as $`$Cachitos.' They arise from the interaction of high-velocity ejecta with an RSG wind, which results in the enhanced excitation of the outer layers of unshocked ejecta and the emergence of corresponding HV absorption \citep{chugai07}. 
The HV components of H\,$\alpha$ and H\,$\beta$ include shallow absorption lines (marked as $`$A' in Figure~\ref{HVplot}) at earlier times ($21-47$~d), and the cachitos appear close to H\,$\alpha$ (marked as $`$B') at later times ($51-71$~d). They are similar to those seen in SN\,2007X \citep[see Figure 13 of][]{gutierrez17}.
\citet{teja23sn23ixf} and \citet{singh24sn23ixfpre}, using a different set of data, report high-velocity absorption features in SN\,2023ixf from 16 d to 70 d after the explosion. \citet{singh24sn23ixfpre} interpret the cachitos as originating from ejecta at different velocities, due to the aspherical distribution of CSM rather than from excitation of unshocked outer ejecta \citep{davis19}.

\subsection{Fitting CO features of SN\,2023ixf}

CO first overtone band emission was detected at eight consecutive epochs from 199 d to 307~d (our final observation) as shown in Figure \ref{nirspec}. The last non-detection of CO was 71~d. Based on NIR observation of SN\,2017eaw \citep{rho18sn} and dust models by \citet{sarangi13}, newly formed CO molecules in SNe II are predicted to be detectable after $\sim$100~d. However, SN\,2023ixf was largely not observable between that day and 199~d because of its proximity to the Sun.  
\new{The first detection of CO emission at 199~d is an upper limit date for the epoch of detection, whereas 71~d is a lower limit. SN\,2017eaw shows a weak CO detection at 107~d with clear detections at 124, 140, 169, and 205~d \citep{rho18sn}. The timing of CO appearance ($>$100~d) is consistent with the chemically controlled dust formation model proposed by \cite{sarangi13}. From the above upper and lower limits, SN\,2023ixf may also have developed CO emission $\sim$100 days since the explosion.}

\begin{figure}[!h]
\centering
\includegraphics[width=\hsize]{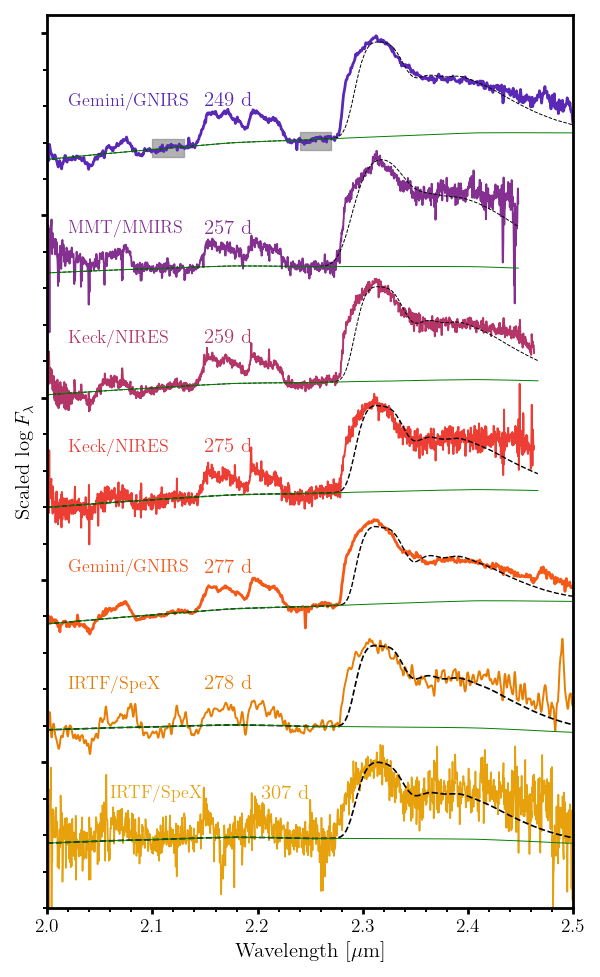}
\caption{Observed Line first-overtone bands in SN\,2023ixf from 249 to 307~d are superposed with the best LTE model fits (black dashed). A linear continuum was adopted for each spectrum. More details are provided in the text. The best-fit parameter values are presented in Table \ref{Tobs}.}
\label{COfittingfull}
\end{figure}

The 2.0--2.5 \mic~portions of the spectra taken between 249 and 307~d are shown in Figure \ref{COfittingfull}. We have modeled the CO emission in them using an LTE model based on \citet{goorvitch94} and \citet{cami10}. \new{We assume pure  $^{12}$C$^{16}$O in our modeling, as previous studies have shown that the CO emission from SNe can be adequately explained by pure $^{12}$CO \citep{banerjee18, rho18sn, rho24}.} 

Our CO model uses the line parameters in \citet{goorvitch94}, which are consistent with those in the HITRAN database{\footnote {https://hitran.org/data-index/}} \citep{rothman05}. In \citet{cami10}, where the molecules are assumed to be isothermal and in LTE, the line strength $S_{\nu_{0}}$ of a line with a transition frequency of $\nu_{0}$ is determined as a function of the CO temperature $T$.  Our model computes the line strength $S_{\nu_{0}}$ following equation (5) of \citet{goorvitch94} and equation (1) of \citet{cami10}:
\begin{equation}
    S_{\nu_0} = \frac{h~\nu_0}{4~\pi}~g_1\,B_{12}~
     \frac{e^{-E_1/kT}}{P(T)} ~
     ({1-e^{-h\nu_{0}/kT})},
   \end{equation}
\begin{equation*} 
    {\hskip 1truecm}= 8.8523\times10^{-13} \frac{gf}{P(T)} \frac{1-e^{-h\nu_{0}/kT}}{e^{E_1/kT}},
\end{equation*}
where $gf$ is the value of the statistical weight $g$ multiplied by the emission oscillator strength $f$, $P(T)$ is the partition function, and $E_1$ is the energy of the lower level. We adopt Einstein-A values, the $gf$ value, and $P(T)$ from \citet[Table 14 of][]{goorvitch94} which lists these values for all the ro-vibration transitions of the first overtone band up to $\upsilon$ = 20 and J = 149.
Each transition experiences broadening following the line profile function $\phi(\nu)$, a Gaussian with the CO velocity $V_{\rm CO}$ as the FWHM. 
\new{We validated that the CO line strengths we calculated match the emission strengths $S$ (column 6) listed in the tables of \cite{goorvitch94} for a temperature of 3000~K.}
The flux of CO emission is calculated as
\begin{equation}
        F_{\nu} = \frac{4 \pi ~(V_{\rm CO}~t)^{2}}{d^2}~B_{\nu}(T_{\rm CO})~\sum_{i} N_{\rm CO}  ~ S^{i}_{\nu_{0}} \phi(\nu,V_{\rm CO})
\end{equation}
where $V_{\rm CO}$ is the velocity of CO, $t$ is the time since the explosion, $d$ is the distance, $N_{\rm CO}$ is the column density of CO, $\phi(\nu)$ is the line profile function, and $B_{\nu}(T)$ is the flux of a blackbody spectrum following the CO temperature $T$ at frequency $\nu$. 
In the fitting process, we use \texttt{kmpfit}, part of the Kapteyn package \citep{KapteynPackage}. The program has been applied to the fundamental CO band of the supernova remnant, Cas A, to reproduce the JWST spectra \citep{rho24} using the mid-infrared CO table \citep[Table 13 from][]{goorvitch94}. \newb{}

The total mass of CO, $M_{\rm CO}$, is related to the velocity and the column density of CO through the following equation:
\begin{equation}
    M_{\rm CO} = 4 \pi~(V_{\rm CO}~t)^{2} ~m~N_{\rm CO}
\end{equation}
where $m$ is the mass of a $^{12}$C$^{16}$O molecule.
We assumed the CO line emission to be optically thin.
This approach is similar to what we used in \citet{banerjee16} and \citet{rho18sn}. 

We compare our LTE model with that of \citet{rho18sn} by applying our model fitting to the spectra of SN\,2017eaw. We find that the CO properties depend on the choice of continuum level. \citet{rho18sn} briefly discussed the idea of different choices of the continuum, but the effects of the different choices were not fully explored. We experimented with 8--10 choices of continuum flux levels for each date and obtained best-fit CO parameters for each (see Figure \ref{fig.appendix1}). We present the median values of CO parameters in the lowermost section of Table \ref{Tobs} with the error values stating the systematic error.
For example, the spectrum at 169~d yields CO temperatures of T$_{\rm CO}$ = ($\sim$ 2750 K, $\sim$ 3050 K, 2915$^{+140}_{-170}$ K), CO velocities of V$_{\rm CO}$ = ($\sim$ 2570, $\sim$ 2890, 2750$^{+135}_{-185}$) \kms, and CO masses of
M$_{\rm CO}$ = ($\sim$ 2.0$\times10^{-4}$, $\sim$1.9$\times10^{-4}$, 1.95$^{+0.09}_{-0.04}$$\times10^{-4}$)\,\msun, for the highest possible continuum flux, the lowest continuum flux, and median values of each 8-10 measurements, respectively. Overall, our LTE model's estimated CO properties are consistent with those of \citet{rho18sn}.

The CO temperatures are determined by the relative strengths of the band heads, resulting in different CO temperatures for different continua. For any given width of CO band heads, a lower CO velocity comes with a lower CO temperature. We also find that the CO mass increases with increasing continuum level, which might be counterintuitive since the CO emission is actually weaker. However, for given strengths of CO emission, a higher CO temperature requires less CO to match the observed band strength. We estimated the uncertainties in CO temperature, velocity, and mass for each choice of continuum. The estimated CO properties of SN\,2017eaw using our LTE model are consistent with those of \citet{rho18sn}.

Following the above exercise, we fit the CO model to each spectrum of SN\,2023ixf using the wavelength ranges $2.10-2.13$ \mic~and $2.24-2.27$ \mic~with the dust emission model to define the flattened continua. We discuss the dust emission model in Section \ref{sec:dustmodel}. The CO emission in each spectrum was fitted over the wavelength range of $2.27-2.46$ \mic, which approximately covers $\upsilon=2-0$ to $6-4$ band heads. 
The estimated CO and dust parameters are shown in Table \ref{Tobs}. We also experimented using linear continua ($`$Model A' in Table \ref{Tobs}) pinned to the spectrum at $2.10-2.11$ \mic~and $2.23-2.25$ \mic\ and using a modified black body continuum
with carbon dust ($`$Model B'); the results were more or less the same for the two models.
In Figure \ref{COfittingfull}, we show the observed and best-fit model spectra for seven of the eight epochs when CO was detected. We could not find reliable CO parameters for the 199~d spectrum from the model fitting because the observation did not cover enough of the CO emission. The CO temperature is constrained by the relative strengths of the different CO bands and requires data at wavelengths longer than 2.36~\mic, which was the long wavelength limit for the spectrum at 199~d.

%Figure4

The results of the CO model fitting between 199~d and 307~d are listed in Table \ref{Tobs}. Based on the model fits, the CO mass, temperature, and velocity were roughly constant from 249 to 278~d: $T_{\rm CO}\sim1800-2300$~K, $V_{\rm CO}\sim 3000-3500$ \kms\ within the errors.
\new{We note that in addition to the standard statistical errors provided in the CO temperature column, the 10-20\% uncertainty in the flux value 
and the telluric correction may result in uncertainties of 100-200 K in the CO temperature and a factor of $\sim$2 in the CO mass estimates from the fitting process.}
%\newb{The CO temperature at 307~d is higher than those at other epochs, and the CO mass at 307~d is lower than those at other epochs. In both Model A and Model B, the temperature difference between 307~d and other epochs is larger than the statistical or systematic uncertainties. However, in Figure \ref{COfittingfull}, we find the CO emission feature at 307~d having two distinct slopes at wavelength ranges $\lambda<2.35$~\mic~and $\lambda>2.35$~\mic, unlike the best-fit model found with the fitting process and similar to CO emission features at other epochs. Considering the small S/N ratio and large uncertainties of the 307~d spectrum, we explore the possibility that the real CO temperature at 307~d is close to the estimates from other epochs and fit the CO emission while fixing the CO temperature to $T=2100$~K for Model A and $T=2250$~K for Model B, which are the estimated temperatures for 278~d. We find CO masses of $M_{\rm CO}\sim0.50\times10^{-4}$~\msun~for Model A and $M_{\rm CO}\sim0.39\times10^{-4}$~\msun~for Model B, larger than the original estimate but smaller than the CO mass found at other epochs.}
%The CO mass was constant to within a factor of $\sim$2 until 278~d. %after which it decreased by an order of magnitude between then and our final measurement at 307~d. 
%A similarly rapid decrease in CO mass was also observed in SN\,1987A between 284 and 349 d in the LTE models of \citet{spyromilio88} as well as in the models of \citet{liu92}. 

%\newb{Overall, we find a trend in the CO mass estimate decreasing with time.}
We find that the estimated CO mass decreases with time.
The model by \citet{liu92} predicts such a decrease because although the formation rate scales as t$^{-3}$, the CO destruction rate scales as $e^{-t/111.3 {\rm d}}$, the decay rate of $^{56}$Co. At later times, cooling of the ejecta also decreases the CO formation rate. In contrast, \cite{sarangi13} predict that total CO mass increases over time, but the CO mass of each individual nucleosynthetic zone could increase or decrease as a function of time.

In Table \ref{Tobs}, we list values of our best-fit CO parameters for spectra acquired using MMT with parentheses to indicate their high uncertainties. 
The MMT spectrum at 257~d extends to 2.45 \mic\ using the K3000 filter.
We find the shape of the MMT spectrum in the K3000 filter range (1.90-2.45 \mic) differs from those of Gemini and Keck spectra at approximately the same date.
Since the K3000 filter is not well used and tested, the systematic errors, including detector transmission and calibration errors, are likely larger than they are for spectra taken at other observatories.

At the bottom of Table \ref{Tobs}, we include the CO parameters estimated from five nebular phase spectra of SN\,2017eaw taken with GNIRS. Spectra taken at $124-205$~d were previously presented in \citet{rho18sn}, but we re-fit them using the procedure described above (see Figure \ref{fig.appendix1}). 

NEOWISE-R detected SN\,2023ixf at $4-11$~d, $211-219$~d, and $370-372$~d \citep{vandyk24sn23ixf}.
At the later dates, a NEOWISE-R excess with a dust temperature of 700~K was reported, which could arise either from newly formed dust or pre-existing dust.  The CO detection indicates that CO cooling occurs in the gas, which can lead to dust formation in the ejecta. Therefore, we suggest that the observed excess continuum (from the black-body function) is from newly formed dust. Unfortunately, the dates do not overlap with our NIR spectra and do not allow us to estimate dust masses by combining our data with the NEOWISE-R coverage.

\subsection{CO emission of SN\,2023ixf and other Type II SNe}
\label{sec:COcomparison}

All of our best-fit models underestimate the CO fluxes in the wavelength intervals $2.28-2.31$~\mic~and $2.41-2.50$~\mic. Similar discrepancies between LTE models and the observed strength of the CO emission at $2.28-2.31$~\mic~are also found in SN\,2017eaw  \citep[see Figure 2 of][]{rho18sn} and 
Type II-pec SN\,1987A \citep[see Figure 1a of][]{spyromilio88}, but not in Type IIb/Ib SN\,2016adj \citep{banerjee18} and Type Ic SN\,2020oi \citep{rho21}. The shorter wavelength interval corresponds to the CO emission near the first ($2-0$) band head, which could be strengthened by considering non-thermal excitation effects \citep{liu92,rho21}. Including CO$^{+}$, whose $(2-0)$ band head is at 2.26~\mic, also increases the flux of the $2-0$ band head as can be seen in the \citet{spyromilio88} spectra of SN\,1987A, but we do not find CO$^{+}$ emission at 2.26 \mic~in our spectra, nor was it present in SN\,2017eaw \citep{rho18sn}. 

The CO emission models exhibit flux declines steeper than the observed spectra at wavelengths $\lambda>2.41$ \mic. The spectrum at 249 d fits better than the others. The same type of discrepancy is also present in LTE models for SN\,1987A at day 349 \citep[see Figure 1 of][]{liu92} and may be present earlier than 349 d (255 or 283 d) when the spectra did not cover wavelengths at $\lambda>2.4$ \mic. This discrepancy is not found in the CO LTE models of \citet{rho18sn} or \citet{banerjee16}, and the difference may be due to them having higher temperatures ($\sim$3000 K) than those of SN\,2023ixf or SN\,1987A ($\sim$2000 K). An alternative explanation could be due to an additional component of dust emission cooler than the component modeled for this fitting.

\citet{liu92} also consider a non-LTE case where the models are closer fits to the observations in the longer wavelength range and yield CO masses about one order of magnitude larger than those of LTE models \citep[see also][]{cherchneff09}.

%Figure5
\begin{figure}
\centering
\includegraphics[width=\hsize]{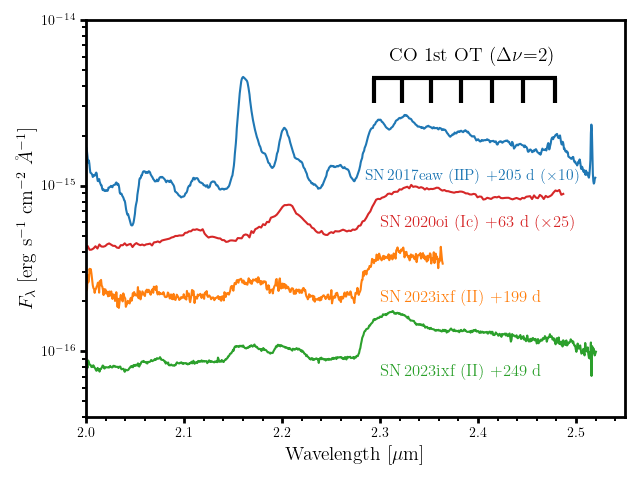}
\caption{The spectra of SN 2023ixf at 199 d and 249 d are compared with those of Type IIP SN 2017eaw at 205 d and Type Ic SN 2020oi at 63 d. All spectra are shifted back to $z=0$, and the latter two are multiplied with constants for visual comparison. Rest wavelengths of the CO first overtone emission band heads for 2-0, 3-1, 4-2, 5-3, 6-4, 7-5, and 8-6 transitions are 2.294, 2.323, 2.353, 2.383, 2.414, 2.446, and 2.479 \mic, respectively (marked as thick black lines).}
\label{NIRcomp}
\end{figure}

We compare the CO features in SN\,2023ixf with those of SN\,2017eaw and SN\,2020oi in Figure \ref{NIRcomp}. The 205 d spectrum of SN\,2017eaw shows clearly distinguished peaks at 2.29 \mic~and 2.32 \mic, corresponding to the $\upsilon=2-0$ and $3-1$ band heads, respectively. The band heads at longer wavelengths are weaker but still identifiable. 
In contrast, the CO emission profiles of SN\,2023ixf and SN\,2020oi are smoother than that of SN\,2017eaw. The 199 d spectrum of SN\,2023ixf and 63d spectrum of SN\,2020oi do not show noticeable peaks at all, while the 249 d spectrum of SN\,2023ixf CO emission rising at 2.28 \mic~to a peak at 2.32 \mic~and connecting to flat emission starting at about 2.35 \mic. The 2.32 \mic~emission bump is broader than the peaks seen from SN\,2017eaw. In the models, smoother CO emission profiles result from a combination of higher CO velocities and lower CO temperatures. For the best-fit parameters of the 249~d CO emission of SN\,2023ixf, we find a similar CO velocity but a lower CO temperature than those for the 205 d CO emission of SN\,2017eaw. For the 63~d spectrum of SN\,2020oi, the CO velocity estimated by \citet{rho21} is slightly higher ($v\sim3700\pm100$ \kms) than the CO velocities of SN\,2023ixf in this work. The 199 d CO spectrum differs from the 249~d spectrum of SN\,2023ixf in that the flux stays flat at $\lambda=2.3-2.35$ \mic, close to the 63 d spectrum of SN\,2020oi. This could be hinting at a higher CO temperature and velocity, but the wavelength coverage is too limited to allow detailed comparison with the models.

The existence of peaks at each CO band head is critical for constraining CO velocity in model fits; their absence complicates the velocity analysis.
In such cases, the emission can be fitted using a LTE model with a larger velocity width. The models show slight bumps at 2.364 \mic~and 2.389 \mic~while the observed spectra do not (e.g., at 277~d). This may imply higher velocities than our estimates.

Comparing the CO emission of SN\,2023ixf with that of other SNe II, we find that the CO temperature for SN\,2017eaw at 205~d ($T_{\rm CO}\sim2400$ K) is similar to that of SN\,2023ixf at 199~d ($T_{\rm CO}\sim2500$ K), although the latter could not be well-constrained due to the reduced wavelength coverage of the MMT data. 
Although we find lower CO temperatures at later times, it is possible that CO emission originate from gas at more than two temperatures, as seen in the young supernova remnant, Cassiopeia A (Cas A) \citep{rho24}. 

%Figure6
\begin{figure}
\centering
\includegraphics[width=\hsize]{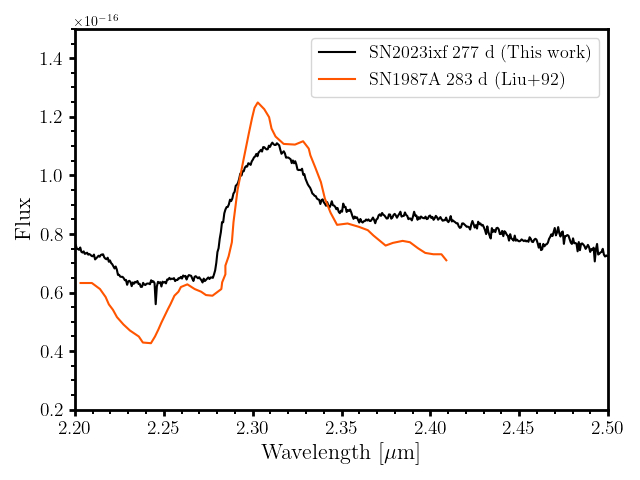}
\caption{The spectrum of SN 2023ixf at 277 d  compared with that of Type II-pec SN\,1987A at 283 d. SN\,2023ixf shows a sharp rise at 2.29~\mic~due to the CO R branch's $\upsilon=2-0$ R branch, and additional emission extending at least to 2.45~\mic,~due to lines of both the CO R and P branches. In contrast, SN\,1987A shows structures near 2.30~\mic and 2.33~\mic~due to the CO $2-0$ and $3-1$ band heads with extended emission at longer wavelength due to lines from both the CO R and P branches.}
\label{87Acomp}
\end{figure}

We also compare the spectra and CO properties of SN\,2023ixf with those of SN\,1987A, as shown in Figure \ref{87Acomp}. The first overtone CO features were detected at phases $250-280$~d in SN\,2023ixf and in SN\,1987A \citep{spyromilio88,liu92}. SN\,2023ixf shows a sharp cut-on caused by the $\upsilon=2-0$ band head (rest wavelength $\lambda=2.2935$~\mic), a broad peak at 2.312~\mic~due to $\upsilon=2-0$ and $3-1$ R branch emission lines, and weaker broad emission staring at $\sim2.36$~\mic~and extending at least to 2.45~\mic~due to a combination of R and P branch lines from a range of vibrational upper levels. In contrast, SN\,1987A shows emission near the $2-0$ and $3-1$ band heads peaking at 2.30 and 2.33~\mic. 
The temperature of SN\,2023ixf is $\sim$2000 ($1600-2500$), while 
\citet{spyromilio88} estimated 
the CO temperature of SN\,1987A at $250-280$~d after the explosion to be $1600-1800$~K \citep{spyromilio88} and $1700-2000$~K \citep{liu92} assuming LTE in the ejecta.
\citet{liu92} also estimated a CO temperature of $2500-2800$~K  using non-LTE assumption.

Our estimates for the CO temperatures \new{($T=1600-2500$ K)} of SN\,2023ixf at 250--280 d are comparable to the LTE estimates for SN\,1987A during the same period. However, unlike the spectra of SN\,1987A which showed two distinct band head peaks at 2.30 and 2.33 \mic, the spectra of SN\,2023ixf show a sole peak at $\lambda\sim2.31$ \mic~due to a larger ejecta velocity. The CO velocity of SN\,2023ixf is 3000--3800 \kms\ while that of SN\,1987A is $\sim$1800 \kms. The LTE models for a CO velocity of 3000 \kms\ with various CO temperatures are shown in Figure \ref{fig.appendix2}. At low temperatures ($T_{\rm CO}<$ 2300 K), the first two band head peaks merge together and appear to be one wide peak.

In addition to the CO velocities of SN\,2023ixf being higher than those of another SN II, we find that the hydrogen lines of SN\,2023ixf are much broader than those of SN\,2017eaw. The FWHM of Pa\,$\beta$ of SN\,2023ixf at 199 d is $\sim$5800 \kms, almost three timers larger than that of SN\,2017eaw at 205 d (FWHM$\sim$2100 \kms).
We suggest that the slanted, boxy profiles observed in the optical spectra of SN\,2023ixf originate from the interaction between the ejecta and the asymmetric CSM \citep{kumar25,bostroem23}. A massive progenitor mass for SN\,2023ixf ($M_{\rm ZAMS}\gtrsim17$ \msun; \citealt{hsu24}) is comparable to that of the luminous, short-plateau SN\,2023ufx ($M_{\rm ZAMS}\sim19-25$ \msun) found by \citet{ravi25} with a notable similarity of shape in boxy Pa\,$\beta$ profiles, but the lines are much broader in SN\,2023ufx than in SN\,2023ixf.

The light curve of SN\,1987A differs from that of a typical SN II, with its luminosity rising until it reaches the peak at $\sim$100~d. The progenitor of SN\,1987A was a blue supergiant \citep{arnett89,dessart19}. On the other hand, the light curve of SN\,2023ixf exhibits a steeper decline than an average SN II during the plateau phase ($\sim$80~d), and its progenitor was a RSG \citep{hosseinzadeh23,bersten24}. We find different CO parameters for SN\,2023ixf and SN\,1987A around the same phase, which might imply the dependency of the onset of CO and dust formation on their progenitors, explosion energies, \nifs\ masses, or SN asymmetries. Observations of additional SNe II are required to discriminate between these possible explanations.

\subsection{Dust Emission from SN\,2023ixf} \label{sec:dustmodel}

The continuum below 1.5 \mic\ can be fit by a black-body function with a temperature of $\sim$5000 K, but the continuum shows excess above 1.5 \mic.
Therefore, we model the dust emission that flattens the spectra at $\lambda>1.5$ \mic, similar to the methods used in \citet{rho21}. The dust continuum was fit with a Planck function [$B_{\nu}(T)$] multiplied by the absorption efficiency ($Q_{abs}$), where we assume carbon dust since it condenses at higher temperatures ($T=1100-1700$~K) than other types of dust \citep{fedkin10}. Details of the absorption efficiency are given in \citet{rho18sn}. 
The dust continuum models for each spectrum are shown in Figure \ref{COfittingfull} as green solid lines.

The dust temperature and mass for each fit from 199 to 307~d are listed in Table \ref{Tobs}. 
We find dust temperatures of $\sim900-1050$~K and dust masses of $\sim(0.7-2.3)\times10^{-5}$~\msun\ from 199 to 307~d. 
\new{In the NIR ($<2.5$~\mic) range, 
we find no clear trends with time in the dust temperature, considering a large systematic error due to the choice of continuum (see Table 1).
The changes in the dust mass of such high dust temperature may indicate the existence of a cooler component.
The dust observed in the NIR must be warm to be observable, and it is likely that additional cooler dust is present, which is undetectable in the NIR and whose mass is expected to increase with time \citep{gall14,wesson21}. Thus, the warm dust ($\sim$1000 K dust in Table 1) inferred from the near-infrared may not accurately estimate the total dust mass. If cooler dust is present, the dust mass can be much greater \citep[examples can be seen in][]{rho18,priestley22}. The warm component is interpreted to be from dust in diffuse ejecta gas \citep{priestley19}.
Indeed, our quick analysis of archival JWST data at day 252 covering both NIR and mid-IR ranges shows significant mid-IR flux excess that supports the existence of a cooler dust component during this period. 
The full JWST analysis is beyond the scope of this paper; a detailed presentation of the mid-IR dust component, along with the complete analysis of the JWST data, will be provided in future studies (\citealp{medlerk25sn23ixf}; W.\, Jacobson-Galán et al., in prep.).}  \new{Several scenarios to explain the flattened continuum in the NIR are discussed in \citet{rho21}. Although a thermal echo from the CSM or heated pre-existing CSM dust cannot be completely ruled out, the simultaneous appearance of warm dust and CO supports the interpretation of dust formation in the ejecta, as CO is known to be a dominant coolant.}
IR spectra with broader and longer wavelength coverage, such as that provided by the JWST and future far-IR telescopes like the Origins Space Telescope, would be needed to accurately estimate the dust temperatures and masses as a function of time.

\section{Summary}

We present a time series of 16 near-infrared spectra of SN\,2023ixf from 9 to 307 d, taken with multiple instruments: Gemini/GNIRS, Keck/NIRES, IRTF/SpeX, and MMT/MMIRS. We quantitatively compared their spectra and found their calibration and spectra are consistent overall. MMIRS made the first CO detection of SN\,2023ixf at 199 d, and Keck/NIRES provided the highest spectral resolution. Gemini/GNIRS covered the longest wavelengths, important for study of the CO first overtone band, and IRTF/SpeX performed well for bright targets such as SN\,2023ixf. The spectra presented are valuable for combining and comparing with the JWST spectra and other ground-based spectroscopic data.

1. In the earliest eight epochs ($t=9-71$~d), narrow (FWHM $\sim100$ \kms) and intermediate-width (FWHM $\sim1000$ \kms) line emission and absorption features originating from the ejecta-wind interaction are present. We found intermediate-width features superimposed on broad P Cygni features during the first 40 days after the explosion. These could originate in the cool, dense shell formed by CSM swept by the shock or from the radiatively accelerated dense CSM near the progenitor. We also found narrow features near the line center of He\one~1.083~\mic~at  47, 67, and 71~d. The intermediate-width and narrow features in NIR lines persist longer than their optical counterparts, which mostly disappeared by 18~d. The temporal evolution of interaction features continues a few tens of days after the explosion; this indicates that it is crucial to view CSM as a product of a complex mass-loss history of the progenitor.

2. We found high-velocity ($\sim$5000--15000~\kms) absorption features (a.k.a. $`$cachitos') in H\,$\alpha$ and H\,$\beta$ from 36~d to 71~d after the explosion. These high-velocity features imply interaction between high-velocity ejecta and an RSG wind.

3. In the final eight epochs ($t=199-307$ d), the continuum is flattened at $1.5-2.5$ \mic~due to early dust formation, which accompanied CO first overtone emission at $\lambda=2.3-2.5$ \mic. Using a LTE model for the CO emission and a modified black body emission for the dust continuum, we fit the CO emission and estimate the CO temperature, CO velocity, CO mass, dust temperature, and dust mass for each epoch. We compare our models with various models including those with non-LTE populations of neutral and singly ionized CO and discuss the discrepancies between our best-fit models and observation. 
\new{We also compare the CO properties of SN\,2023ixf with those of other Type II supernovae, SN\,2017eaw and SN\,1987A, at similar evolutionary phases. We find that while the CO temperatures of SN\,2023ixf at $250-300$~d are comparable to those of SN\,2017eaw at similar evolutionary phases, the CO velocities are much higher, suggesting diversity in CO and dust formation.}
%We also compare the CO temperatures of SN\,2023ixf with those of other Type II supernovae, SN\,2017eaw and SN\,1987A, at similar evolutionary phases and find that SN\,2023ixf has lower CO temperatures at $250-300$ d than SN\,1987A, implying a variety in the CO and dust formation. 

We found that models with only neutral $^{12}$C$^{16}$O were able to reproduce the CO spectra of SN\,2023ixf.
Previous model NIR spectra of other SNe \citep[e.g.,][]{liu92,rho21} that included non-LTE energy distributions and CO$^{+}$ ions were able to reproduce observed CO$^{+}$ emission at shorter ($\lambda\lesssim2.3$~\mic) wavelengths. We discuss the limitations of the single dust component assumption in our models. 
Further studies, including non-thermal processes, are encouraged to gain a better understanding of the phenomenon.

\begin{acknowledgements}
Based on observations obtained at the international Gemini Observatory, a program of NSF NOIRLab, which is managed by the Association of Universities for Research in Astronomy (AURA) under a cooperative agreement with the U.S. National Science Foundation on behalf of the Gemini Observatory partnership: the U.S. National Science Foundation (United States), National Research Council (Canada), Agencia Nacional de Investigaci\'{o}n y Desarrollo (Chile), Ministerio de Ciencia, Tecnolog\'{i}a e Innovaci\'{o}n (Argentina), Minist\'{e}rio da Ci\^{e}ncia, Tecnologia, Inova\c{c}\~{o}es e Comunica\c{c}\~{o}es (Brazil), and Korea Astronomy and Space Science Institute (Republic of Korea). We thank Gemini staff %, including Ho-Gyu Lee, Sang-Hyun Chun, Monika Soraisam, and Zachary Hartman, 
for supporting the Gemini GNIRS observations.
The University of Arizona team would like to thank J. Hinz, G. Williams, and the rest of the MMT staff for enabling and supporting the MMIRS observations reported here.
This work makes use of observations from the Las Cumbres Observatory network.
S.-H. P. and S.-C.Y. were supported by the National Research Foundation of Korea (NRF) NRF-2019R1A2C2010885 and NRF-2022H1D3A2A01096434. S.-C. Y. was supported by the NRF RS-2024-00356267.
J.R. was partially supported by a NASA ADAP grant (80NSSC23K0749), JWST-GO-01947.032, and Brain Pool visiting program for Outstanding Overseas Researchers by NRF-2022H1D3A2A01096434.
The LCO team is supported by NSF grants AST-1911225 and AST-1911151. Time-domain research by the University of Arizona team and D.J.S.\ is supported by National Science Foundation (NSF) grants 2108032, 2308181, 2407566, and 2432036 and the Heising-Simons Foundation under grant \#2020-1864.  C.L. acknowledges support from the National Science Foundation Graduate Research Fellowship under Grant No. DGE-2233066 and DOE award DE-SC0010008 to Rutgers University.
K.A.B is supported by an LSST-DA Catalyst Fellowship; this publication was thus made possible through the support of Grant 62192 from the John Templeton Foundation to LSST-DA.
L.G. acknowledges financial support from AGAUR, CSIC, MCIN and AEI 10.13039/501100011033 under projects PID2023-151307NB-I00, PIE 20215AT016, CEX2020-001058-M, and 2021-SGR-01270.
Observations reported here were obtained at the MMT Observatory, a joint facility of the University of Arizona and the Smithsonian Institution.
Observations reported here were taken at the Infrared Telescope Facility, which is operated by the University of Hawaii under contract 80HQTR24DA010 with the National Aeronautics and Space Administration.
This research is based in part on observations obtained at the international Gemini Observatory, 
a program of NSF’s NOIRLab, which is managed by the Association of Universities for Research in Astronomy (AURA) under a cooperative agreement with the National Science Foundation on behalf of the Gemini Observatory partnership: the National Science Foundation (United States), National Research Council (Canada), Agencia Nacional de Investigación y Desarrollo (Chile), Ministerio de Ciencia, Tecnología e Innovación (Argentina), Ministério da Ciência, Tecnologia, Inovações e Comunicações (Brazil), and Korea Astronomy and Space Science Institute (Republic of Korea).
Some of the data presented herein were obtained at Keck Observatory, which is a private 501(c)3 non-profit organization operated as a scientific partnership among the California Institute of Technology, the University of California, and the National Aeronautics and Space Administration. The Observatory was made possible by the generous financial support of the W. M. Keck Foundation. 
The authors wish to recognize and acknowledge the very significant cultural role and reverence that the summit of Maunakea has always had within the Native Hawaiian community. We are most fortunate to have the opportunity to conduct observations from this mountain. 
\end{acknowledgements}

\section*{Software}

Astropy \citep{astropy:2013, astropy:2018, astropy:2022}, numpy \citep{harris2020array}, scipy \citep{2020SciPy-NMeth}, \texttt{XDGNIRS} \citep{xdgnirs}, FLOYDS pipeline \citep{valenti14}, MMIRS pipeline \citep{mmirspipe}, XTELLCOR \citep{cushing04}, Spextool \citep{cushing04}, \texttt{pypeit} pipeline \citep{pypeit2020,prochaska2020}, YSE-PZ \citep{2022zndo...7278430C,2023PASP..135f4501C}, \texttt{kmpfit} \citep{KapteynPackage}, matplotlib \citep{Hunter:2007} 
    
\bibliography{msrefsallsn23ixf}

\begin{appendix}
\section{Additional Figures}
\begin{figure}[!hb]
\centering
\includegraphics[width=8cm]{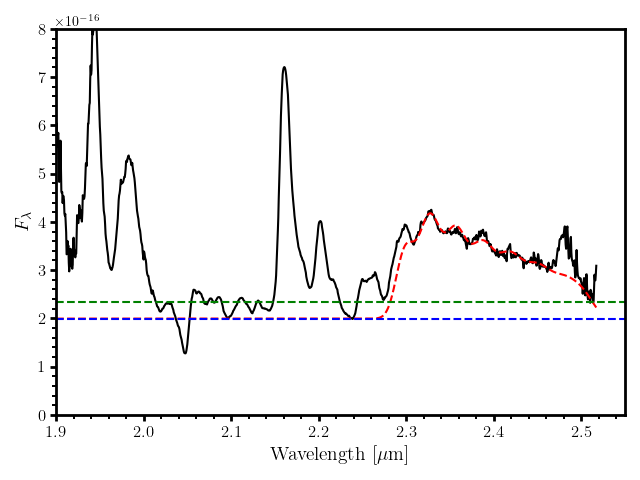}
\caption{CO first overtone emission of SN 2017eaw at 169 d since the explosion compared to the LTE emission model described in the text (red dotted). We demonstrate the possibility of different continua with a lower limit (blue dotted) and an upper limit (green dotted) of the continuum. 
}
\label{fig.appendix1}
\end{figure}

\begin{figure}[!hb]
\centering
\includegraphics[width=8.5cm,height=11truecm]{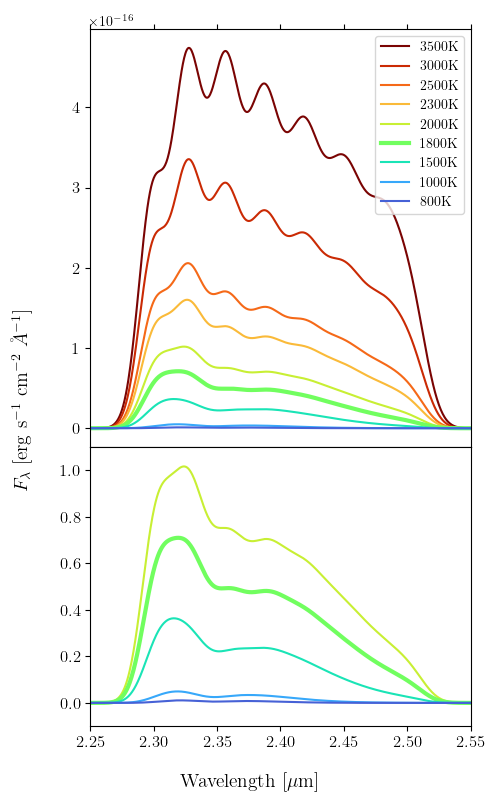}
\caption{CO LTE model spectra for a given velocity of 3000 \kms\ and a range of temperatures between 800 and 3500 K. Higher CO temperature shows clearer band heads. Lack of distinct spectral features due to band heads from $\upsilon=3$ and higher vibrational levels indicates the CO temperature $<$ 2300 K.
}
\label{fig.appendix2}
\end{figure}

\end{appendix}

\end{document}